\documentclass[twocolumn,aps,prb]{revtex4}
\usepackage{graphicx}
\newcommand{\be}{\begin{equation}}
\newcommand{\ee}{\end{equation}}

\newcommand{\bea}{\begin{eqnarray}}
\newcommand{\eea}{\end{eqnarray}}
\newcommand{\bd}{\begin{displaymath}}
\newcommand{\ed}{\end{displaymath}}
\newcommand{\bi}{\begin{itemize}}
\newcommand{\ei}{\end{itemize}}
\newcommand{\bc}{\begin{center}}
\newcommand{\ec}{\end{center}}
\newcommand{\bfl}{\begin{flushleft}}
\newcommand{\efl}{\end{flushleft}}
\newcommand{\bfr}{\begin{flushright}}
\newcommand{\efr}{\end{flushright}}
\newcommand{\f}{\frac}

\def\bk{{\bf k}}

\def\6{\partial}  
  \def\ve{\varepsilon}

  \def\D{\Delta}

\def\={\!\!\!&=&\!\!\!}
\def\+{\!\!\!&&\!\!\!+~}
\def\-{\!\!\!&&\!\!\!-~}

\begin{document}
\date{\today}

\title{Field-induced Bose-Einstein condensation of interacting
dilute magnons in three-dimensional spin systems: A
renormalization-group study}

\author
{M. Crisan$^1$, I. \c{T}ifrea$^{1,2}$, D Bodea$^{1,3}$, and I.
Grosu$^1$}

\affiliation{$^1$Department of Physics, ``Babe\c{s}-Bolyai"
University, 40084 Cluj Napoca, Romania\\ $^2$Department of Physics
and Astronomy, University of Iowa, Iowa City, IA 52242,
USA\\$^3$Max Planck Institute for the Physics of Complex Systems,
01187 Dresden, Germany}

\begin{abstract}
We use the Renormalization Group method to study the magnetic
field influence on the Bose-Einstein condensation of interacting
dilute magnons in three dimensional spin systems. We first
considered a model with $SU(2)$ symmetry (universality class
$z=1$) and we obtain for the critical magnetic field a power law
dependence on the critical temperature, $[H_c(T)-H_c(0)]\sim T^2$.
In the case of $U(1)$ symmetry (universality class $z=2$) the
dependence is different, and the magnetic critical field depends
linearly on the critical temperature, $[H_c(T)-H_c(0)]\sim T$. By
considering a more relevant model, which includes also the
system's anisotropy, we obtain for the same symmetry class a
$T^{3/2}$ dependence of the magnetic critical field on the
critical temperature. We discuss these theoretical predictions of
the renormalization group in connection with experimental results
reported in the literature.
\end{abstract}

\maketitle

\section{Introduction}

Bose--Einstein condensation (BEC) remains one of the most
important macroscopic effects predicted long time ago by quantum
mechanics for an ideal Bose gas. Lately, the interest in BEC was
renewed by the practical realization of a condensate phase in weak
interacting Bose systems realized in ultra--cooled diluted alkali
atomic gases \cite{BECalkali}. Further, a BEC of fermionic pairs
was achieved in trapped fermionic systems under attractive
fermion--fermion interaction \cite{BECFermi}. The limitation of
these experiments is due to a reduced number of particles which
undergo BEC. Recently, BEC was associated with the magnetic
transition observed in different quantum spin systems. A
particular class of such materials is formed by the XCuCl$_3$
dimer compounds (with X being Tl, K, or NH$_4$) \cite{ta,oo}. In
the ground state these materials are spin singlets which are
magnetically inactive. If a high--magnetic field is applied, they
undergo a transition into a magnetically ordered state, a
transition which can be understood as a condensation of
excitations which behaves as bosonic quasiparticles.

One particular example is TlCuCl$_3$. The quantum magnetism in
this compound is attributed to the spin 1/2 Cu$^{2+}$ ions
positioned in the double chains of Cu$_2$Cl$_6$. The ground state
of TlCuCl$_3$ was found to be a singlet with an excitation gap
$\D\simeq 7.5$ K \cite{shiramura}. The gap has been associated
with the weak anisotropic antiferromagnetic (AF) intra--dimer
coupling in the double chain. Magnetic susceptibility experiments
at different temperatures for different directions of the applied
magnetic field exhibit broad maxima at $T=38$  K and a decreasing
behavior toward zero susceptibility with decreasing temperature
\cite{oo}. At very low temperatures, the magnetic susceptibility
for different values of the applied magnetic field behaves quite
differently. At fields of the order of 1 T, the magnetic
susceptibility decreases exponentially to zero, a proof for the
existence of a ground state gap. On the other hand, for a magnetic
field of 7 T an anomaly in the magnetic susceptibility was
reported around $T=4$ K. This is an indication that the ground
state must be gapless at high magnetic fields. The system has a
three dimensional character and in an external magnetic field $H$,
the singlet--triplet gap $\Delta$ is reduced to $\Delta-\mu_B g
H$, and vanishes at a critical field $H_c=\Delta/\mu_B g$.
Inelastic neutron scattering measurements proved the existence of
the elementary magnonic excitations with a strong dispersion in
all three directions \cite{ca1,oo2}. The observed gap,
$\Delta=0.7$ meV, which is much smaller than the intra--dimer
interaction $J=5$ meV, and the small critical field $H_c= 5.6$ T,
make this material a perfect candidate to study the magnetic field
induced  phase transition. Another argument in favor of the
possible phase transition was the evidence for a Goldstone mode
\cite{ru} observed also by inelastic neutron scattering
experiments \cite{ca2}. Neutron diffraction experiments at fields
$H>H_c$ showed that the field induced AF order in the plane normal
to the applied field appears at the same time with the uniform
moments \cite{ta2}. Sound attenuation experiments were associated
to a ``relativistic"--like form of the excitations energy
spectrum, $E(k)=\sqrt{\Delta^2+k^2/[2m_{eff}]}$ ($m_{eff}=2/J^2$),
a relation which was used to explain the connection between the
transition temperature and the concentration of magnons \cite{sh}.
The main conclusion of all these experimental approaches is that
TlCuCl$_3$ undergoes a magnetic ordering in high magnetic fields.
The temperature dependence of the critical magnetic field can be
summarized by $[H_c(T)-H_c(0)]\propto T^{\phi}$, where from the
experimental data fit $\phi\simeq 1.7-2.2$.

Various theoretical models were used to investigate the BEC of
magnons in quantum spin liquids. In the mean field approximation
most of the experimental features of the phase transition cannot
be reproduced, in particular, the critical temperature dependence
on the magnetic field being almost flat \cite{tachikiMFT}. The
possibility of a magnetic field induced phase transition in
quantum spin liquids was first discussed in terms of BEC by
Giamarchi and Tsvelik \cite{gi}. This possibility was investigated
using the Hartree--Fock approximation introduced by Popov
\cite{popov}. The initial system is mapped into a dilute Bose
liquid with a magnetic field dependent chemical potential
$\mu=g\mu_B(H-H_c)$, and the total number of magnons $N$ is
associated with the total magnetization $M$, $M=g\mu_B N$
\cite{ni}. The theory predicts $[H_c(T)-H_c(0)] \sim T^{3/2}$, a
result which is not in complete agreement with experimental data.
The model was further on investigated by Misguich and Oshikawa
\cite{misu} by including a more realistic dispersion relation
\cite{ca1} and the shape of the critical temperature curves as
function of the critical magnetic field was well reproduced with a
single adjustable parameter.

The possible phase transition was numerically investigated by
Monte Carlo simulation \cite{no1,ka,no2}. Nohadani {\it et al.}
\cite{no1} found that the value of the critical temperature
exponent $\phi$ is in general greater that the Hartree--Fock value
$\phi=1.5$, but is close to it as the magnetic field approaches
its critical zero temperature value. Kawashima \cite{ka} argued
that the Hartree--Fock value $\phi=1.5$ may be incorrect at finite
temperature, but is the correct value in the $T=0$ K limit, namely
for the situation when the phase transition is actually a quantum
phase transition (QPT). The BEC of magnons as a QPT at $T=0$ K was
further investigated by Monte Carlo simulation by Nohadani {\it et
al.} \cite{no2}. The model considered a control parameter
$g=J'/J$, where $J$ is the intra--dimer exchange interaction, and
$J'$ is the inter--dimer coupling. The experimental data gives for
these couplings the values $J=60-70$ K and $J'=40-53$ K. However,
those values can be controled by external pressure. The resulting
$T=0$ K phase diagram consists  of three different regions,
regulated by the $g$ ratio. At low magnetic fields and small
control parameter $g$ the system is a dimer spin liquid, i.e., is
in a magnetically disordered state. This phase is characterized by
a $SU(2)$ symmetry with a dynamical critical exponent $z=1$. At
intermediate magnetic fields $H<H_s$ ($H_s$ is the saturation
field) and sufficiently large value of $g$, the ground state is
partially spin polarized for $H>H_c$ ($H_c$ is the critical field)
and has a long--range antiferromagnetic order transverse to the
applied magnetic field direction. In this case the phase symmetry
is only $U(1)$ with a corresponding dynamical critical exponent
$z=2$, a value associated to the quadratic dispersion of the
bosonic excitations. At large fields $H>H_{s}$ ($H_s \sim J+5J'$),
all spins are fully polarized.

Here we consider a Renormalization Group (RG) approach of the QPT
associated to the BEC of magnons in spin dimer systems. It is well
known that the presence of a QPT will influence the system's
properties even at finite temperatures. The exact temperature
range in the vicinity of the $T=0$ K phase transition is unknown,
however, as in any phase transition a critical region will be
present. We will consider separately the phases identified by
Monte Carlo simulation, and show that different values for the
critical temperature exponent $\phi$ are obtained in the RG
analysis. We will show that in the phase with a $SU(2)$ symmetry
$\phi=2$, whereas for the phase with $U(1)$ symmetry $\phi=1$. If
one considers a more realistic model with a $U(1)$ symmetry and
strong anisotropy, $\phi=3/2$. This behavior suggests the
existence of a narrow crossover temperature interval close to
$T=0$ K ($T\ll1$ K) where the critical temperature coefficient
complies with the Hartree--Fock value $\phi=3/2$. At higher
temperature, still in the critical region, the symmetry of the
system changes and the value of the temperature critical exponent
is higher $\phi\simeq 2$. We will discuss those results in
connection with the numerical calculations and experimental data.

\section{Renormalization Group Analysis}

Following Hertz \cite{he} and  Millis  \cite{mi}, we describe the
critical behavior of a magnon dilute gas in terms of an effective
Ginzburg--Landau--Wilson theory of the order parameter field
$\Phi(\bf k,\tau)$ which represents the fluctuations of the
staggered magnetization of the system.

In the general theory of QPT's the total action will consist on
two terms, $S[{\Phi}]=S_2[\Phi]+S_4[\Phi]$. The quadratic part,
$S_2[\Phi]$, takes the form
\begin{equation}
S_2[\Phi]=\frac{1}{2}\sum_k\Phi^\dagger(k)[r_0(H)+k^\sigma
+(i\omega_n)^m]\Phi(k)\;,
\end{equation}
where $k\equiv({\bf k}, \;{\omega_n})$  with ${\omega_n}=2n{\pi
T}$ ($k_B=1$), $\Phi(k)$ is a bosonic field describing the
magnetization fluctuations, and $r_0(H)$ measures the distance
from the quantum critical point. In the presence of an external
magnetic field the control parameter $r_0(H)$ will acquire a field
dependence. The form of the action was extensively discussed by
Fisher and Rosch \cite{ro} both on a phenomenological basis and
starting from a Hubbard--type model of electrons moving in the
presence of a magnetic field. In the following we will consider to
have a Zeeman type $H$ dependence of the control parameter,
$r_0(H)\sim [H-H_c]$. The dimensionality of the system is $d=3$,
and $\sigma$ and $m$ will take different values according to the
studied phase, i.e., $\sigma=2$ and $m=2$ meaning that the
dynamical critical exponent $z=1$ for the $SU(2)$ symmetry, and
$\sigma=2$ and $m=1$ with a dynamical critical exponent $z=2$ for
the $U(1)$ symmetry. The interacting contribution to the action,
$S_4[\Phi]$, will be of standard form
\begin{eqnarray}
&&S_4[\Phi]=\nonumber\\ &&\frac{u_0}{16} \sum_{k_1,\ldots,k_4}
\Phi(k_1) \Phi(k_2)\Phi^\dagger(k_3)\Phi^\dagger(k_4) \delta
(k_1+k_2+k_3+k_4)\;,\nonumber\\
\end{eqnarray}
and describes interactions between the considered fluctuations.
The total action will remain invariant under a standard
transformation of momenta, frequency, and fields, i.e., $k'=kb$,
$\omega_n'=\omega_n b^z$, where $z$ is the dynamical critical
exponent, and $\Phi'= \Phi b^{-(d+z+2)/2}\;$. \cite{sa}

The scaling equations for the parameters $T$, $r$, and $u$ have
the general form \cite{il,cri}:
\begin{eqnarray}
\label{eqT}&&\frac{dT(l)}{dl}=zT(l)\;,\\
%
\label{eqr}&&\frac{dr(l)}{dl}=2r(l)+K_3 F_1[r(l),T(l)]\;,\\
%
\label{equ}&&\frac{du}{dl}=[4-(d+z)]u(l)-\frac{5}{2}K_3F_2[r(l),T(l)]u^2(l)\;,\;
\end{eqnarray}
where $F_1[r(l),T(l)]$ and $F_2[r(l),T(l)]$ are complicated
functions of parameters $T$ and $r$. Their values in the low
temperature domain and at the critical point ($r=0$) are given in
Ref. \cite{il} as $F_1[T(l)]=\coth{[1/(2T(l))]}/2$ and
$F_2[T(l)]=1/4$. Also, $K_3$ is a constant whose value is
$1/(2\pi^2)$. Equations (\ref{eqT})--(\ref{equ}) will be solved to
analyse the system behavior in the critical region for two
situations. First we will analyse the $SU(2)$ symmetry situation
when the dynamical critical exponent $z=1$, and then the $U(1)$
symmetry case when $z=2$.

\subsection{SU(2) symmetry}

One possible phase identified by Monte Carlo simulations was
characterized by the limit of low magnetic fields and small
coupling ratios $J'/J$, being a magnetically disorder phase, i.e.,
a dimer spin liquid. In this case the dynamical critical
coefficient value is $z=1$. Two of the renormalization group
equations, namely (\ref{eqT}) and (\ref{equ}), can be solved
exactly with the following solutions
\begin{equation}
T(l)=T\;e^l
\end{equation}
and
\begin{equation}\label{solu}
u(l)=\frac{1}{C_0(l+l_0)}\;,
\end{equation}
where $C_0=5K_3/8 $  and  $l_0=1/[C_0 u_0]$. The remaining
equation (\ref{eqr}), the one for the control parameter $r$, will
be solved considering a solution of the form $r(l)\simeq
r_0\exp(2l)h(l)$, leading to
\begin{equation}\label{solr}
r(l)=e^{2l}\left[r_0+\frac{K_3}{4}u_0+K_3I_r^{(3)}(l)\right]
 -\frac{K_3u(l)}{4}\;,
\end{equation}
where
\begin{equation} I_r^{(3)}(l)=\int _0
^{l}\frac{dx\exp[-2x]u(x)}{\exp{[1/T(x)]}-1}\;.
\end{equation}
To study the influence of the criticality on the thermodynamics of
the $SU(2)$ phase we introduce the scaling field $t_r(l)$ defined
as:
\begin{equation}
t_r(l)=r(l)+\frac{K_3}{4}u(l)\;,
\end{equation}
a field which will be used to stop the renormalization process.
Based on Eqs. (\ref{solu}) and (\ref{solr}) we get
\begin{equation}
t_r(l)=e^{2l}\left[t_r(0)+K_3\zeta(2)u_0T^2\right]\;,
\end{equation}
where $t_r(0)=r_0-r_{0c}$ with $r_{0c}=-K_3u_0/4$, and $\zeta(x)$
is the Riemann zeta function. The renormalization procedure will
be stopped at $l=l^*\gg 1$, given by $t_r(l^{\star})=1$. To find
$l^*$ we rewrite $t_r(l)=e^{2l}t_r(T)$ and after simple
calculations one finds
\begin{equation}
l^{\star }\sim \ln\frac{1}{T}\;.
\end{equation}
We can define a critical line whose equation is given by
$t_r(T)=0$ which permits us to calculate the critical field
$H_c(T)$ as:
\begin{equation}
H_c(T)=H_c(0)-K_3u_0 \zeta(2)T^2\;.
\end{equation}
Accordingly, in the $SU(2)$ symmetry phase we can identify the
exponent of the critical temperature dependence on the magnetic
field to be $\phi=2$, a value which is close to the one reported
in the literature.

\subsection{U(1)symmetry}

Another possible phase identified by Monte Carlo simulations of
the quantum phase transition in spin dimer liquids has $U(1)$
symmetry and is present for intermediate applied magnetic fields
and a sufficiently large value of the ratio $J'/J$. In this case
the dynamical critical exponent value is $z=2$, and the
renormalization group equations (\ref{eqT})--(\ref{equ}) are
similar to those of the weakly interacting Bose gas
\cite{fisher,cri}. Equation (\ref{eqT}) has a trivial solution of
the form
\begin{equation}
T(l)=Te^{2l}\;.
\end{equation}
We consider now equation (\ref{equ}) which gives us the
renormalization of the interaction parameter. If we introduce
$\ve=2-d<0$ ($d=3$) this equation can be rewritten in the form
\begin{equation}
\frac{du(l)}{dl}=\ve u(l)-\frac{K_3}{4} u^2(l)\;,
\end{equation}
and admits the following solution
\begin{equation}\label{usolU1}
u(l)=\frac{u_0\exp[-|\varepsilon|l]}{1+K_3\left(\exp[-|\varepsilon|l]-1\right)/4}\;.
\end{equation}
One can proceed now to the solution for the renormalized control
parameter $r$. In the $U(1)$ symmetry case this equation will
admit the following solution
\begin{eqnarray}\label{eqrlU1}
&&r(l)\nonumber\\&&=e^{2l}\left\{r_0+\frac{K_3u_0}{4}+
\frac{K_3Tu(l)}{2}\ln\left[\f{1}{1-\exp[-1/T(l)]}\right]\right\}\nonumber\\
&&-\frac{K_3u(l)}{4}\;.
\end{eqnarray}
As in the previous situation we introduce the scaling field
$t_r(l)=r(l)+K_3u(l)/4$ and we will stop the renormalization
procedure at $t_r(l^*)=1$. With this definition, based on equation
(\ref{eqrlU1}), the general form of the scaling field become
\begin{equation}
t_r(l)=e^{2l}\left\{t_r(0)+\frac{K_3u(l)}{2}T\ln{\left[\f{1}{1-\exp[-1/T(l)]}\right]}\right\}\;,
\end{equation}
where $t_r(0)=r_0-r_{0c}$ with $r_{0c}=-K_3u_0/4$. We stop the
scaling procedure at $l=l^*\gg 1$, where once again $l^*$ is the
solution of $t_r(l^\star)=1$. In this case the equation is more
complicated, however, in the low temperature regime we can
approximate its solution. If we rewrite $t_r(l)=e^{2l}t_r(T)$ with
\begin{equation}
t_r(T)\simeq r_0-r_{0c}+\f{K_3T}{2}u_0\ln\frac{1}{u_0}\;,
\end{equation}
we finally find
\begin{equation}
l^*\sim \f{1}{2}\ln{\f{1}{T}}\;.
\end{equation}
The critical line, identified from $t_r(T)=0$ will give us the
critical magnetic field $H_c(T)$ as function of temperature
\begin{equation}
H_c(T)=H_c(0)-\left[\frac{u_0}{4\pi^2}\ln\frac{1}{u_0}\right]T.
\end{equation}
In the case of $U(1)$ symmetry the relation between the critical
field and temperature is different from the case of $SU(2)$
symmetry and we identified in this situation the critical
temperature exponent to be $\phi=1$.

\section{Influence of anisotropy}

According to Monte Carlo calculations at sufficiently large fields
the system has a ground state which is partially spin polarized
and has antiferromagnetic long-range order transverse to the
magnetic field direction. The fact that the magnetic properties of
the system depend on direction suggests that anisotropy may play
an important role when the system is in this partially polarized
phase. To take in account such effects, we consider that the
interaction term in the total action corresponding to the $U(1)$
symmetry phase has the following form \cite{ka,no2}:
\begin{eqnarray}
S_4[\Phi]&=&\frac{1}{16} \sum_{k_1,\ldots,k_4}\Phi(k_1)
\Phi(k_2)\Phi^\dagger(k_3)\Phi^\dagger(k_4) \delta_{\bk_1+\bk_2;\bk_3+\bk_4}\nonumber\\
&&\times\left[u_0\delta_{\omega_{n1}+\omega_{n2};\omega_{n3}+\omega_{n4}}+
v_0\delta_{\omega_{n1};\omega_{n3}}\delta_{\omega_{n2};\omega_{n4}}\right]\;,\nonumber\\
\end{eqnarray}
where the coupling constants $u_0$ and $v_0$ describe interactions
between the magnetization fluctuations in different directions of
the system. Using the same procedure as in the previous cases the
renormalization group equations can be obtained as
\begin{eqnarray}\label{eqTan}
&&\frac{dT(l)}{dl}=2T(l),\\
&&\frac{dr(l)}{dl}=2r(l)\nonumber\\
&&\hspace{1cm}+\frac{K_3}{2}\left\{[v(l)+2u(l)]F_1[T(l)]+v(l)T(l)\right\},\;\;\\
&&\frac{du(l)}{dl}\simeq\varepsilon u(l)-\frac{K_3}{4}u^2(l)\;,\\
\label{ecvan} &&\frac{dv(l)}{dl}\simeq \varepsilon v(l)\;.
\end{eqnarray}
The last two equations, the ones corresponding to the renormalized
coupling constants $u(l)$ and $v(l)$, have been written in the
lowest order, an approximation which is assumed valid in the low
temperature domain. Equation (\ref{eqTan}) admits the trivial
solution $T(l)=Te^{2l}$.  The first coupling constant, $u(l)$, has
the same form no matter if one considers or not the anisotropy,
given by equation (\ref{usolU1}). Equation (\ref{ecvan}) admits
the trivial solution $v(l)=v_0e^{-|\varepsilon| l}$. The solution
for the remaining equation which is giving us the renormalized
phase transition control parameter, $r$, can be written as
\begin{eqnarray}
r(l)=e^{2l}&&\left\{r_0-r_{0c}+\frac{K_3}{2}\int_0^l
dl'v(l')T(l')\right.\nonumber\\
&&\left.+\frac{K_3}{2}\int_0^ldl'[v(l')+2u(l')]F_1[T(l')]\right\},\;\;\;
\end{eqnarray}
where $r_{0c}=K_3[v_0+2u_0]/8$. We can introduce again the scaling
field $t_r(l)=r(l)+K_3[v(l)+2u(l)]/8$ which based on the solution
for $r$ can be written as
\begin{eqnarray}\label{tran}
t_r(l)&=&e^{2l}\left\{r_0-r_{0c}+\frac{K_3v_0}{2\varepsilon}T[e^{-|\varepsilon|
l}-1]\right.\nonumber\\
&&\left.+\frac{K_3}{4}T[v(l)+2u(l)]\ln\left[\f{1}{1-\exp[-1/T(l)]}\right]\right\}\;,\nonumber\\
\end{eqnarray}
%
where $\varepsilon<0$ for $d=3$. To calculate the value of the
stop scaling parameter $l^*$ we consider $t_r(l^\star)=1$ and, in
the lowest order, we obtain $l^\star$ as:
\begin{equation}
l^{\star}\sim\f{1}{2}\ln\f{1}{T}.
\end{equation}
Equation (\ref{tran}) for the scaling field can be rewritten in
the form $t_r(l)=e^{2l}t_r(T)$ where $t_r(T)$ can be evaluated as:
\begin{eqnarray}
t_r(T)&=&r_0-r_{0c}+\frac{K_3v_0T}{2|\ve|}\nonumber\\
&&+\frac{K_3}{4}(v_0+2u_0)T^{3/2}\ln\left[\f{1}{v_0+2u_0}\right].
\end{eqnarray}
In the anisotropic $U(1)$ case the temperature dependence of the
critical magnetic field will be calculated in the limit $u_0>v_0$
from $t_r(T)=0$ as
\begin{equation}\label{fieldan}
H_c(T)\simeq
H_c(0)-\left[\frac{u_0}{2\pi^2}\ln\frac{1}{2u_0}\right]T^{3/2}\;.
\end{equation}
A similar result was predicted using different calculation methods
and numerical evaluation in Refs. \cite{ka,no2}. Equation
(\ref{fieldan}) predicts an important role for the anisotropy in
the case of $U(1)$ symmetry, as the critical temperature exponent
changes from $\phi=1$ to $\phi=3/2$. The later value is close to
the lowest observed experimental value, i.e. 1.7, and identical to
the value reported by Monte Carlo studies.

\section{Discussions}

The idea of BEC in solid state systems was associated with
electronic Cooper pairs in superconductors, excitons in
semiconductors, and more recently with magnons in spin liquid
dimer compounds. The occurrence of a condensate phase in these
compounds was investigated by various theoretical methods, from
mean field to renormalization group approximations. In the case of
spin liquid dimer compounds the BEC of magnons is induced by
magnetic fields, the critical field associated to the BEC being
temperature dependent and characterized by a critical exponent
$\phi$, defined as $[H_c(T)-H_c(0)]\sim T^\phi$. The exact value
of the critical exponent $\phi$ can vary from 1.7 to 2.2 according
to experimental data. A mean field analysis gives $\phi=1.5$, and
for some time it was believed that such approximation was enough
to explain all relevant physics of the phase transition.

Here we applied the renormalization group method to study the
possible BEC of magnons. We started our analysis from the premise
that a quantum phase transition, despite being characterized by a
$T=0$ K critical temperature, will influence the system properties
even at finite temperatures. Accordingly, we analyze the phase
transition in the vicinity of a quantum phase transition, in the
low temperature limit. The possible quantum phase transition in a
spin liquid dimer was investigated by Monte Carlo simulations
considering a model which includes also the system's anisotropy
\cite{no2}. As function of anisotropy the phase diagram of a spin
liquid dimer will consists of three different regions, in each of
them the system being in a different symmetry class. \cite{no2} At
low magnetic fields, the system is in a magnetically disordered
state. As the fields increases, a partially spin polarized state
will develop and the symmetry of the system will change
accordingly. At high magnetic fields the system is in a
long--range order antiferromagnetic phase. To take into account
all these possibilities, we analyzed the influence of magnetic
fields and temperature on the BEC of magnons in systems with
different symmetry. In the case of a magnetically disordered spin
liquid dimer, when the symmetry of the system is in the $SU(2)$
class and the dynamical critical exponent is $z=1$, the critical
exponent is $\phi=2$. On the other hand, when the partially
ordered phase is induced by the magnetic field, the system's
symmetry changes to $U(1)$, and the dynamical critical exponent
becomes $z=2$. In this situation, we calculated $\phi=1$, a value
which was never observed in real experiments. However, in this
partially ordered state anisotropy plays an important role, as the
phase is characterized by a large $J'/J$ ratio. When anisotropy is
taken into account by considering direction dependent interactions
in the action, the value of the critical coefficient $\phi$
changes from 1 to 1.5, a value which is close to the lowest
reported experimental data. The influence of anisotropy was
considered by Fischer and Rosh \cite{ro} including additional
terms in the free action, the final result for the temperature
critical exponent being $\phi=1.5$, the same value we obtained.
The crossover problem was also considered in a sigma model
framework to explain some phases of the organic insulator
(TMTTF)$_2$PF$_6$ as function of temperature, magnetic field, and
pressure. \cite{brown,schmeltzer} A similar idea was used to
explain nuclear magnetic resonance experimental data in
superconducting spin--ladder compounds such as
Sr$_2$Ca$_{12}$Cu$_{24}$O$_{41}$. \cite{bishop}

In conclusion, we showed that the renormalization group analysis
of the possible BEC of magnons in spin liquid dimers is a more
appropriate investigation method. This method can account for the
change in the critical exponent $\phi$ from 1.5 to 2, according to
the phase symmetry. The results are in good agreement with the
experimental data in TlCuCl$_3$ samples. More recently a similar
behavior was reported also in Cs$_2$CuCl$_4$ samples \cite{teo}.
Such a change in the critical exponent cannot be explained by mean
field approximations. On the other hand the ground state of
TlCuCl$_3$ is strongly influenced by pressure.\cite{johannsen}
Accurate experimental data obtained for various values of the
applied external pressure still support previous results for the
critical exponent, namely $\phi=2.6$. \cite{johannsen} Finally, we
mention that recent experimental data showed that the occurrence
of the BEC of magnons can be also induced by hydrostatic pressure
\cite{uw}. The dependence of the critical field on pressure, and
the pressure dependence of the transition temperature are
suggesting a similar theoretical description.

\end{document}